\documentclass[aps,pra,superscriptaddress,amsmath,twocolumn,amssymb]{revtex4}

\usepackage{xr}

\usepackage{subfigure}
\usepackage{amsmath}
\usepackage{pgfplots}
\pgfplotsset{compat=1.5}
\usepgfplotslibrary{groupplots}

\usepackage{color}

\usepackage{xcolor}
\usepackage{bm}
\usepackage{multirow}
\usepackage{placeins}
\usepackage{soul}
\usepackage{color,xcolor}
\usepackage[colorlinks,linkcolor=red,anchorcolor=blue,citecolor=blue,urlcolor=blue]{hyperref}
\usepackage{cleveref}

\newcommand{\Hop}{\hat{H}}
\newcommand{\sgx}{\hat{\sigma}^x}
\newcommand{\sgy}{\hat{\sigma}^y}
\newcommand{\sgz}{\hat{\sigma}^z}
\newcommand{\circuit}{\mathcal{C}}
\newcommand{\fidelity}{\mathcal{F}}
\newcommand{\rhoop}{\hat{\rho}}
\newcommand{\Iop}{\hat{I}}
\newcommand{\tr}{{\rm tr}}
\newcommand{\para}{\vec{\theta}}
\newcommand{\conj}{{\rm conj}}
\newcommand{\bdim}{\chi}
\newcommand{\cswap}{{\text {c-SWAP}}}

\begin{document}

\title{Variational Quantum Circuits for Quantum State Tomography}

\author{Yong Liu}\thanks{These authors contribute equally to this work.}
\affiliation{Institute for Quantum Information \& State Key Laboratory of High Performance Computing, College of Computer, National University of Defense Technology, Changsha 410073, China}

\author{Dongyang Wang}\thanks{These authors contribute equally to this work.}
\affiliation{Institute for Quantum Information \& State Key Laboratory of High Performance Computing, College of Computer, National University of Defense Technology, Changsha 410073, China}

\author{Shichuan Xue}
\affiliation{Institute for Quantum Information \& State Key Laboratory of High Performance Computing, College of Computer, National University of Defense Technology, Changsha 410073, China}

\author{Anqi Huang}
\affiliation{Institute for Quantum Information \& State Key Laboratory of High Performance Computing, College of Computer, National University of Defense Technology, Changsha 410073, China}

\author{Xiang Fu}
\affiliation{Institute for Quantum Information \& State Key Laboratory of High Performance Computing, College of Computer, National University of Defense Technology, Changsha 410073, China}

\author{Xiaogang Qiang}
\affiliation{Institute for Quantum Information \& State Key Laboratory of High Performance Computing, College of Computer, National University of Defense Technology, Changsha 410073, China}
\affiliation{National Innovation Institute of Defense Technology, AMS, 100071 Beijing, China}

\author{Ping Xu}
\affiliation{Institute for Quantum Information \& State Key Laboratory of High Performance Computing, College of Computer, National University of Defense Technology, Changsha 410073, China}

\author{He-Liang Huang}
\affiliation{Henan Key Laboratory of Quantum Information and Cryptography, IEU, Zhengzhou 450001, China}
\affiliation{Hefei National Laboratory for Physical Sciences at Microscale and Department of Modern Physics,\\
University of Science and Technology of China, Hefei, Anhui 230026, China}
\affiliation{CAS Centre for Excellence and Synergetic Innovation Centre in Quantum Information and Quantum Physics,\\
University of Science and Technology of China, Hefei, Anhui 230026, China}

\author{Mingtang Deng}
\affiliation{Institute for Quantum Information \& State Key Laboratory of High Performance Computing, College of Computer, National University of Defense Technology, Changsha 410073, China}

\author{Chu Guo}
\email{guochu604b@gmail.com}
\affiliation{Quantum Intelligence Lab, Supremacy Future Technologies, Guangzhou 511340, China}

\author{Xuejun Yang}
\affiliation{Institute for Quantum Information \& State Key Laboratory of High Performance Computing, College of Computer, National University of Defense Technology, Changsha 410073, China}

\author{Junjie Wu}
\email{junjiewu@nudt.edu.cn}
\affiliation{Institute for Quantum Information \& State Key Laboratory of High Performance Computing, College of Computer, National University of Defense Technology, Changsha 410073, China}

\begin{abstract}
  Quantum state tomography is a key process in most quantum experiments. In this work, we employ quantum machine learning for state tomography. Given an unknown quantum state, it can be learned by maximizing the fidelity between the output of a variational quantum circuit and this state. The number of parameters of the variational quantum circuit grows linearly with the number of qubits and the circuit depth, so that only polynomial measurements are required, even for highly-entangled states. After that, a subsequent classical circuit simulator is used to transform the information of the target quantum state from the variational quantum circuit into a familiar format. We demonstrate our method by performing numerical simulations for the tomography of the ground state of a one-dimensional quantum spin chain, using a variational quantum circuit simulator. Our method is suitable for near-term quantum computing platforms, and could be used for relatively large-scale quantum state tomography for experimentally relevant quantum states.
\end{abstract}

\date{\today}

\maketitle

\section{Introduction}

Identifying a quantum state is a key step to verify or benchmark any quantum processes~\cite{Smithey1993,Vogel1989,Leonhardt1995,Dunn1995}. Practically, quantum state tomography (QST) is a standard technology to obtain the information of an unknown state through quantum measurements and is widely used in many quantum experiments~\cite{Lu2007,Haffner2005,James2001}. The efficiency of QST highly depends on the times of quantum measurements as well as the copies of the target states.

However, the original technique for QST requires an exponentially growing number of measurements, which could be feasible only for a small number of qubits~\cite{White1999}. To alleviate this exponential scaling in the original proposal, various approaches has been used based on some assumptions about the structure of the target state. An outstanding class of these methods is based on tensor network states~\cite{Cramer2010,Zhao2017,Wang2017}, which can efficiently represent quantum states with bounded entanglement entropy through a polynomial number of parameters~\cite{Lanyon2017}. However, the tensor-network-based methods may not suit for highly entangled states where these methods still suffer from exponential scaling. Other examples along this line include the permutationally invariant tomography~\cite{Toth2010, Moroder2012}, as well as the compressed sensing which reduces the number of measurements with the assumption that the target quantum state is sparse~\cite{GrossEisert2010, LiuYuan2012}. Besides these methods, the classical-neural-network-based methods also obtain remarkable achievements~\cite{Torlai2018, Torlai2018b,Rocchetto2018,Quek2018,Carrasquilla2019}, where the information of the target state is encoded with the parameters of the neural network. These methods all represent the quantum state in classical manners, and have to face the challenge of exponential scaling in specific cases. For example, the matrix product state (MPS) representation becomes inefficient for highly entangled states, such as the output state of a random quantum circuit (RQC)~\cite{BoixoNeven2018}.

Recently, the rapid development of quantum hardware~\cite{Arute2019} indicates that the computing power of quantum processors can be used for specific applications. In this work, we show that by using a quantum processor, it only requires a polynomial number of measurements for quantum state tomography, even for certain highly entangled states. We apply quantum machine learning (QML)~\cite{Biamonte2017}, a possibly suitable application for noisy intermediate-scale quantum (NISQ) computer, to learn the target state and encode its information into the parameters of a variational quantum circuit (VQC). In this way, the target state can be stored within a more natural quantum data structure. The number of parameters of this circuit grows polynomially with the number of qubits. Unlike the tensor-network-based methods, with these polynomial number of parameters, the VQC can approximate highly entangled states. Therefore, our method can serve for a wider range of applications. After encoding the information of the target state into the circuit, an optional step can be to transfer the information of this state into a familiar format (such as a state vector). Actually, this process is done via the classical simulation of the state evolution through this circuit, which is an exponentially hard problem for classical computers. Straightforwardly, one can represent the state by a vector for the simulation, but in this paper, we also introduce a MPS circuit simulator which can be scaled up to cases with more qubits. We demonstrate our method by the numerical simulations of the tomography for the ground state of a one-dimensional quantum spin chain with $6\sim 15$ qubits on a personal computer.

This paper is organized as follows. In Sec.\ref{sec:vqc}, we introduce the scheme of our quantum machine learning algorithm for quantum state tomography. In Sec.\ref{sec:mps}, we show how to extract the information of the target state from the VQC with a MPS circuit simulator. In Sec.\ref{sec:result}, we demonstrate our method with numerical simulations of quantum state tomography for the ground state of a quantum spin chain. Finally, we conclude in Sec.\ref{sec:summary}.

\section{Approximating quantum states with quantum machine learning}\label{sec:vqc}

\begin{figure}[t]
  \centering
  \includegraphics[width=0.48\textwidth]{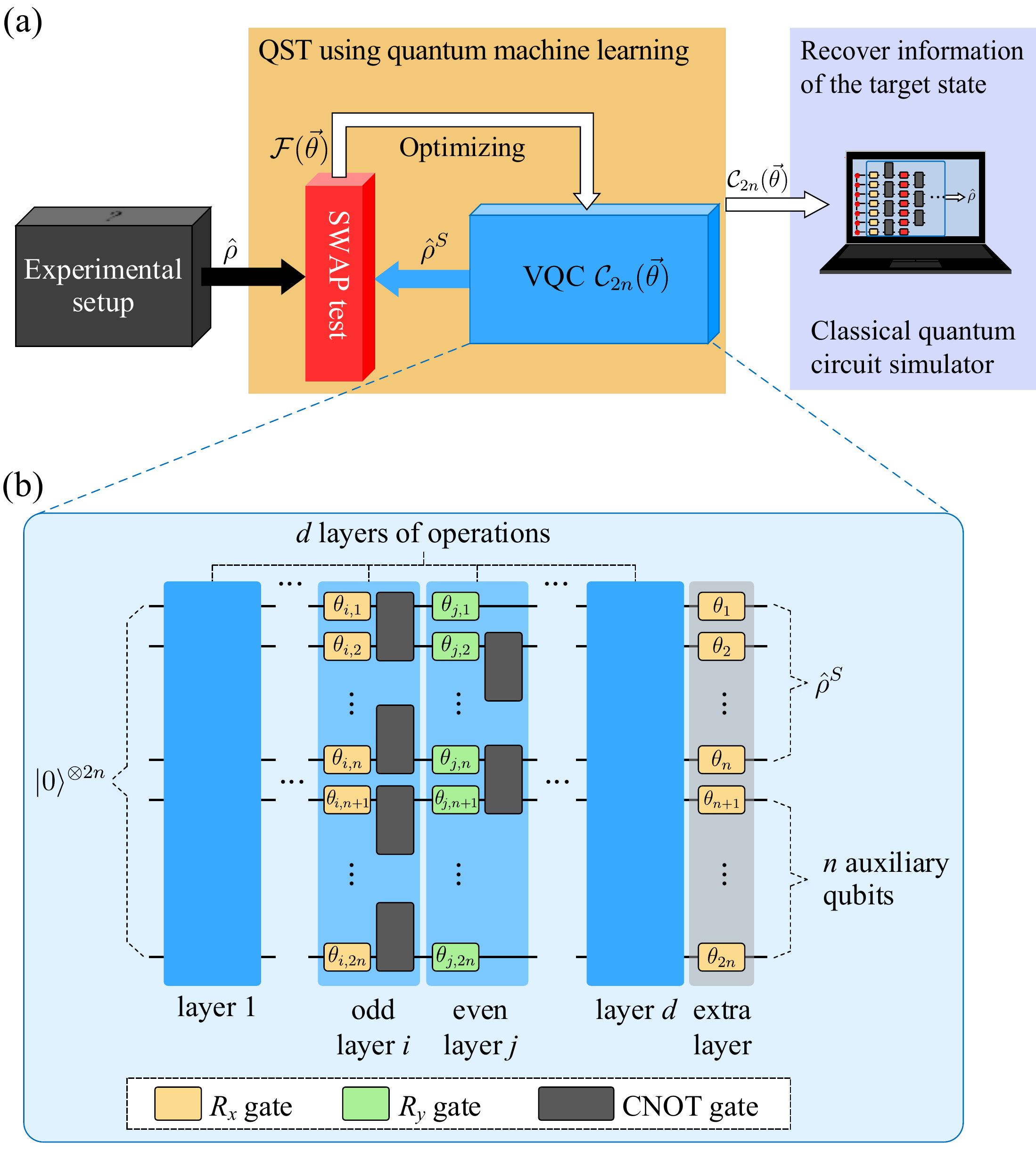}\\
  \caption{(a) The scheme of the approach to quantum state tomography. The information of the unknown state is learned by maximizing the fidelity $\mathcal{F}(\vec{\theta})$ measured through the SWAP test between the reduced density matrix of the first $n$ qubits in the output state $\hat{\rho}^S$ and the unknown state $\hat{\rho}$. The information of the state is then stored with a quantum data structure: the variational circuit with optimized parameters. Then the information of the state can be transferred into a familiar format (e.g. a state vector) using a classical circuit simulator. (b) The structure of the variational quantum circuit for state estimation. The circuit contains $d$ layers of operations. The $i^{th}$ layer contains $2n$ single-qubit gates encoding the tunable parameters $\left\{\theta_{i,1}, \theta_{i,2}, \cdots, \theta_{i,2n}\right\}$, and a group of commuting controlled-not (CNOT) gate operations applied on the neighbouring qubits alternately. The parametric single-qubit gates are respectively $R_x$ gates in the odd layers or $R_y$ gates in the even layers. The circuit ends up with an extra layer of single-qubit operations. The latter $n$ auxiliary qubits are unnecessary if we have the prior knowledge that the unknown state is pure.}\label{FIG:VQC}
\end{figure}

The information of the unknown state is obtained through a quantum machine learning algorithm as shown in Fig.~\ref{FIG:VQC}(a). Generally, the target state under tomography is an $n$-qubit mixed state, which we denote as $\rhoop$. To fully capture the information in $\rhoop$, we use a $2n$-qubit variational quantum circuit $\circuit_{2n}(\para)$ (since it is enough to purify any $n$-qubit mixed state with $n$ auxiliary qubits~\cite{Ping2013}), where $\para$ contains all the parameters to be optimized. The output of the variational quantum circuit is denoted as $\vert\psi^o\rangle$, which can be written as
\begin{align}
\vert\psi^o\rangle = \circuit_{2n}(\para)\vert 0\rangle^{\otimes 2n}.
\end{align}
The reduced density operator of the first $n$ qubits can then be obtained by
\begin{align}
\rhoop^S = \tr_A\left(\vert\psi^o\rangle\langle \psi^o\vert \right),
\end{align}
where $\tr_A$ means the partial trace over the latter $n$ qubits. The fidelity between $\rhoop^S$ and $\rhoop$ can be represented by
\begin{align}
\fidelity(\para) = \tr_S(\rhoop  \rhoop^S) = \langle \psi^o \vert \rhoop\otimes\Iop \vert \psi^o \rangle,
\end{align}
where $\tr_S$ means the trace over the former $n$ qubits.
Note that $\fidelity(\para)$ can be efficiently computed with a quantum computer via SWAP test~\cite{Buhrman2001}.

The goal is then to maximize $\fidelity(\para)$ over $\para$, for which we simply choose the loss function of our quantum machine learning algorithm as
\begin{align}\label{eq:lossopen}
f(\para) = 1 - \sqrt{\fidelity(\para)}.
\end{align}
Specifically, if the target state is known to be a pure state in advance, then only $n$ qubits are required in our variational circuit, i.e. $|\psi^o\rangle=\circuit_n(\para)\vert 0\rangle^{\otimes n}$. In this case, the loss function can be simplified to be
\begin{align}\label{eq:lossunitary}
f(\para) = 1- | \langle \psi^o\vert \psi\rangle |.
\end{align}

In Fig.~\ref{FIG:VQC}(b) we show a possible implementation of the VQC, which consists of interlacing layers of single-qubit rotation gates and two-qubit CNOT gates. To represent generic quantum states, both parametric rotational X ($R_x$) gates and rotational Y ($R_y$) gates are used, which are defined as
\begin{align}
R_x(\theta)&=\left[\begin{matrix}\cos\frac{\theta}{2} & -i\sin\frac{\theta}{2} \\ -i\sin\frac{\theta}{2} & \cos\frac{\theta}{2}\end{matrix}\right],\\
R_y(\theta)&=\left[\begin{matrix}\cos\frac{\theta}{2} & -\sin\frac{\theta}{2} \\ -\sin\frac{\theta}{2} & \cos\frac{\theta}{2}\end{matrix}\right].
\end{align}
In this circuit, each layer of commuting CNOT gates is counted as one depth. Thus, for such a circuit with depth $d$, the total number of parameters is $2n(d+1)$ in general. However, if we have the priori knowledge that the target state is pure, then the number of parameters can be reduced to $n(d+1)$, and the simplified loss function described by Eq.(\ref{eq:lossunitary}) can be used during the optimization. It is worth noting that the structure of the circuit in Fig.~\ref{FIG:VQC}(b) closely resembles that of a random quantum circuit~\cite{Chen2020}, and the random quantum circuit can generate statistic distributions, which are intractable for classical computers to produce~\cite{BoixoNeven2018}.

We apply a gradient-based optimization method to iteratively update the parameters $\para$ and minimize the loss function.
Using the chain rule, we have
\begin{align}\label{eq:grad}
\frac{\partial f(\para)}{\partial\theta_i}=\frac{\partial f(\para)}{\partial \fidelity(\para)} \frac{\partial \fidelity(\para)}{\partial \theta_i}.
\end{align}
The first term on the right hand side of Eq.(\ref{eq:grad}) can be easily computed using a classical computer as long as we have obtained the value of $\fidelity(\para)$ from a quantum computer, and the second term $\partial \fidelity(\para)/\partial \theta_i$ can also be computed on a quantum computer through~\cite{Mitarai2019}
\begin{equation}\label{EQ:G1}
\begin{aligned}
  \frac{\partial \fidelity(\para)}{\partial \theta_i}&=\frac{1}{2}\fidelity(\para_i^+) -\frac{1}{2}\fidelity(\para_i^-),
\end{aligned}
\end{equation}
where $\para_i^{\pm}$ is the array of parameters obtained by adding or subtracting the $i^{th}$ parameters of $\para$ by $\pi/2$.

For one iteration of the algorithm, besides the evaluation of $\fidelity(\para)$, it is also required to evaluate $\fidelity(\para_i^\pm)$ for all $2n(d+1)$ parameters to calculate the gradient. A single evaluation of $\fidelity(\para)$ requires to execute the circuit for a constant number of times to reach a certain precision in the $n$-qubit SWAP test, and each execution involves around $3nd$ gate operations. As a result, the complexity of each iteration is $O(n^3d^2)$.

\section{State reconstruction with MPS}\label{sec:mps}

The VQC will store almost all the information of the target state as long as the above quantum machine learning algorithm manages to minimize the loss function to approximate $0$, which indicates that $\rhoop^S \approx \rhoop$. We can then reconstruct the state by transforming the information encoded in the circuit to a familiar format, namely to simulate this circuit classically. One can directly store the full wavefunction as a state vector, and then simulate the state evolution, which however requires an exponential amount ($O(2^{2n})$) of memory. To accelerate the classical processing, we can apply the MPS algorithm for the information recovery. In the following we show how to reconstruct $\rhoop^S$ with MPS~\cite{McCaskey2018}.
The MPS representation of the $2n$-qubit state can be written as~\cite{Schollwock2011}
\begin{equation}\label{eq:mps}
|\phi\rangle=\sum_{\sigma_1,...,\sigma_{2n}}\mathcal{G}\left(B^{\sigma_1}B^{\sigma_2}\cdots B^{\sigma_{2n}}\right)\vert \sigma_1, \sigma_2, \dots, \sigma_{2n} \rangle .
\end{equation}
Each $B^{\sigma_l}_{a_l,a_{l+1}}$ is a rank-3 tensor, where $\sigma_l$ represents the physical index and $a_l$ represents the auxiliary index. Function $\mathcal{G}$ means summation over common auxiliary indices. Here we also assume that the MPS is prepared in the \textit{right canonical} form, namely $B^{\sigma_l}_{a_l,a_{l+1}}$ satisfies
\begin{align}\label{eq:rightcanonical}
\sum_{\sigma_l, a_{l+1}} B^{\sigma_l}_{a_l,a_{l+1}}\conj(B^{\sigma_l}_{a_l',a_{l+1}}) = \delta_{a_{l}, a_{l}'},
\end{align}
where $\conj(M)$ means to take the elementwise conjugate of the tensor $M$, and $\delta_{i, j}$ is the Kronecker matrix satisfying $\delta_{i, j} = 1$ for $i=j$ or $0$ otherwise. The maximum size of the auxiliary indices is referred as the bond dimension $\bdim$, namely
\begin{align}
\bdim = \max_{1\leq l\leq 2n-1} \dim(a_l).
\end{align}

\begin{figure}[t]
  \centering
  \includegraphics[width=0.48\textwidth]{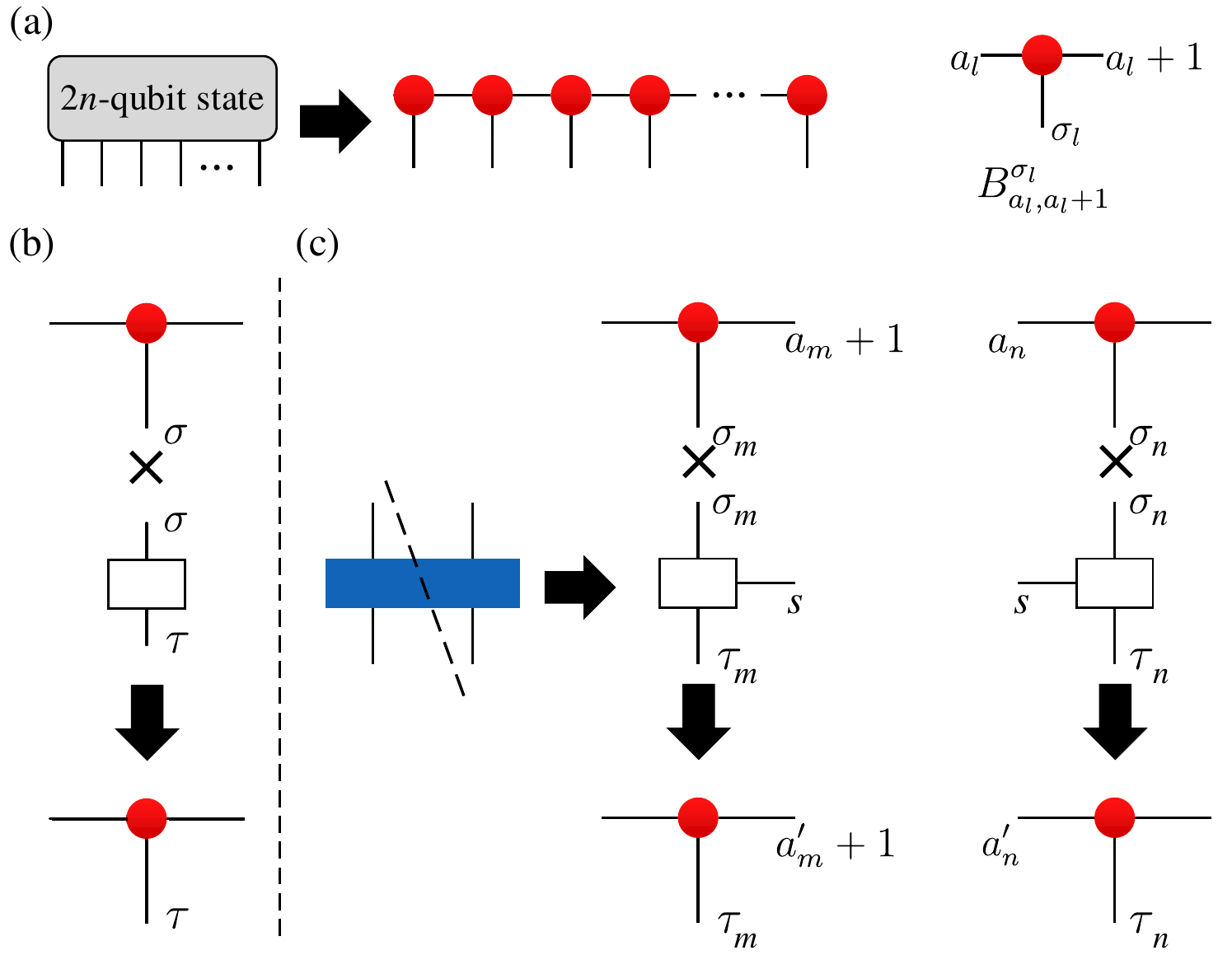}\\
  \caption{The classical circuit simulator based on MPS. (a) A $2n$-qubit quantum states represented with MPS. Each qubit is represented by a rank-3 tensor $B^{\sigma_l}_{a_l,a_l+1}$. (b) Applying a single-qubit operation on a local tensor of MPS, which does not affect the sizes of the tensors. (c) A two-qubit gate operation is first decomposed into two local operations and then applied to nearest-neighbour qubits respectively. The sizes  of the tensors increase in general after this operation.}\label{FIG:OP}
\end{figure}

The initial state of the VQC, $\vert 0\rangle^{\otimes 2n}$, can be easily written as a separable MPS with $\chi=1$, which is shown in Fig.~\ref{FIG:OP}(a). Then the single-qubit and two-qubit gates are applied to this MPS in a way that the right canonical form of the MPS is preserved. The single-qubit and two-qubit gate operations on MPS are shown in Fig.~\ref{FIG:OP}(b, c) respectively. For detailed mathematical description of these operations, one can refer to, for example, Ref.~\cite{Schollwock2011, HuangGuo2019}. Note that each time a two-qubit gate is performed on a pair of neighbour qubits, and the bond dimension will effectively increase by a factor of $\bdim_o$, which is the rank of the two-qubit operation. As to the CNOT gate, we have $\bdim_o = 2$~\cite{GuoLiu2019}. As a result, for a variational quantum circuit organized as in Fig.~\ref{FIG:VQC}(a) with depth $d$, the final MPS will have a bond dimension
\begin{align}
\bdim \leq 2^{\frac{d}{2}}.
\end{align}

After the evolution, we trace out the latter $n$ qubits of the resulting MPS and obtain the reduced density matrix in the form of a matrix product operator (MPO)
\begin{align}\label{eq:rhos}
\rhoop^S = \sum_{\sigma_1, \dots, \sigma_n, \sigma_1', \dots, \sigma_n'}&\mathcal{G}\left(B^{\sigma_1}\dots B^{\sigma_n} B^{\sigma_1'}\dots B^{\sigma_n'}\right) \nonumber \\  &\vert \sigma_1, \dots, \sigma_n \rangle\langle \sigma_1', \dots, \sigma_n' \vert,
\end{align}
where
\begin{align}
&\mathcal{G}\left(B^{\sigma_1}\dots B^{\sigma_n} B^{\sigma_1'}\dots B^{\sigma_n'}\right) \nonumber \\
=& \sum_{\substack{a_1, \dots, a_n, a_{n+1} \\ a_1', \dots, a_n', a_{n+1}}} B^{\sigma_1}_{a_1, a_2}\dots B^{\sigma_n}_{a_n, a_{n+1}} B^{\sigma_1'}_{a_1', a_2'}\dots B^{\sigma_n'}_{a_n', a_{n+1}}.
\end{align}
Note that we have exploited the property of the MPS described by Eq.(\ref{eq:rightcanonical}) during the simulation.

Finally, the obtained MPO contains all the information of the target state, whose size is bounded by $4n\bdim^2 = 4n2^d$. The component of the target state in a particular basis $\vert \tau_1, \dots, \tau_n\rangle\langle \tau_1', \dots, \tau_n'\vert$ can be computed by
\begin{align}
\langle \tau_1', \dots, \tau_n' \vert \rhoop^S \vert \tau_1, \dots, \tau_n\rangle = \mathcal{G}\left(B^{\tau_1}\dots B^{\tau_n} B^{\tau_1'}\dots B^{\tau_n'}\right),
\end{align}
where the computational complexity is $O(\bdim^3)$.

Although here we have proposed to reconstruct the target quantum state as a MPS (or more generally a MPO), we stress that our approach is entirely different from the approaches in Ref.~\cite{Cramer2010,Lanyon2017}. The reasons are as follows.
\begin{itemize}
\item [i)] The efficiency of the quantum state tomography, which highly depends on the times of measurement and copies of target states, does not necessarily rely on the assumption that the target quantum state has a limited amount of entanglement.
\item [ii)] When recover the information of the target state using the MPS circuit simulator, the bond dimension $\chi$ of the MPS representation will in general grow exponentially. Therefore the main focus of our approach is to reduce the number of quantum gate operations or the number of quantum measurements.
\end{itemize}

A straightforward example to show the differences can be the output state of the one-dimensional random quantum circuit~\cite{Chen2020}. The MPS tomography may require exponential measurement because of the rapidly growing entanglement, while in our method, only polynomial measurements are required because the structure of the VQC and RQC can be the same. Though the subsequent classical MPS simulator has exponential complexity,  it is still tolerable with a moderate classical computer for mixed states with less than $20$ qubits or pure states with less than $40$ qubits. Actually, our method is similar with a recent work that applies a parametric Hamiltonian~\cite{XinLi2020} which, however, is only suitable for pure states.

\section{Numerical simulation and performance analysis}\label{sec:result}

We demonstrate our method by numerical simulations based on a VQC simulator~\cite{Farhi2018,Kandala2017}. Although our method can be used to approximate density operators in general, here we consider the case where the target state is pure. Moreover, it is the ground state of a local spin Hamiltonian, the Heisenberg XXZ spin chain
\begin{align}
\Hop_{XXZ} =& \sum_{l=1}^{L-1}\left[J\left(\sgx_{l}\sgx_{l+1} + \sgy_l\sgy_{l+1} \right) +  \Delta \sgz_l\sgz_{l+1}\right] \nonumber \\ &+ h \sum_{l=1}^{L} \sgz_l,
\end{align}
where $L$ is the number of spins (qubits), $h$ is the magnetization strength, $J$ is the tunneling strength, and $\Delta$ is the interaction strength. In the simulations, we fix $h=1$ and $J=1$, and $L$ is fixed according to the manner of fidelity calculation. Therefore, the ground state can be treated as a function of $\Delta$, which we denote as $\vert GS(\Delta)\rangle$. $\Hop_{XXZ}$ is gapless when $\Delta \leq 1$ or gapped when $\Delta > 1$, and it is worth noting that the ground state of a gapped Hamiltonian can be efficiently computed either with a classical computer~\cite{Landau2015} or a quantum computer~\cite{Bilgin2010}. With the priori knowledge that the `unknown' quantum state is pure, we can simply use the loss function defined in Eq.(\ref{eq:lossunitary}). Moreover, since $\vert GS(\Delta)\rangle$ contains only real numbers, we only use parametric $R_y$ gates in our circuits.
\begin{figure}[t]
  \centering
  \begin{tabular}{cc}
  \includegraphics[width=0.24\textwidth]{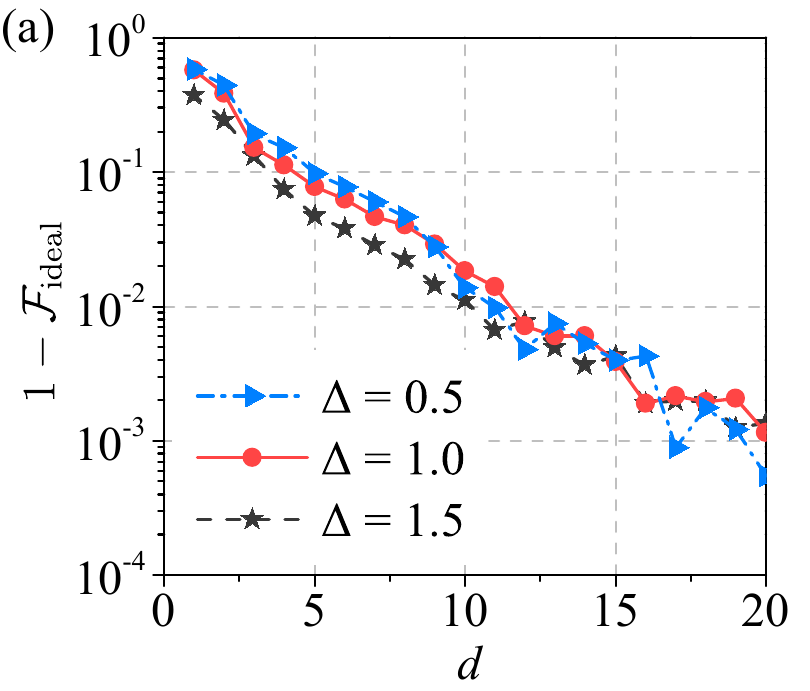}&
  \includegraphics[width=0.24\textwidth]{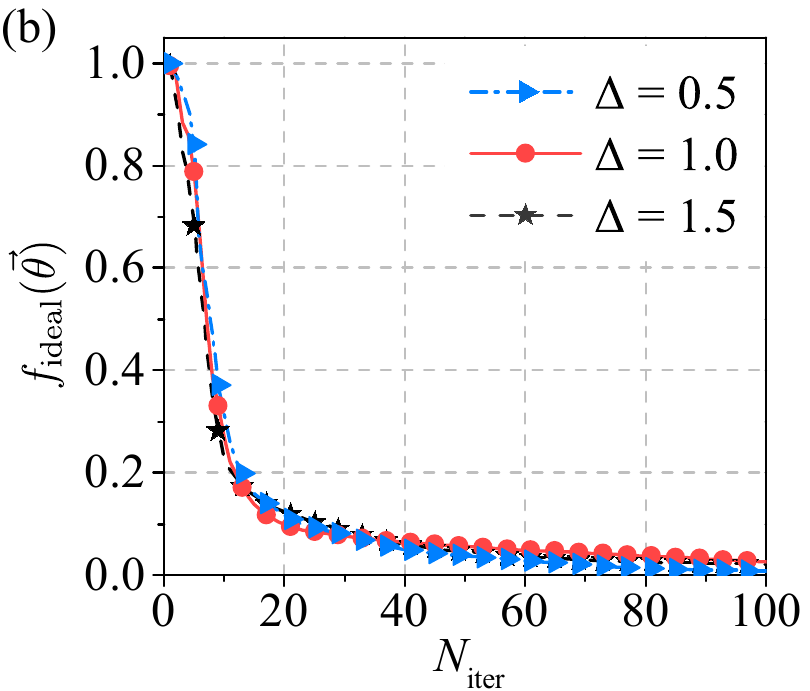}\\
  \includegraphics[width=0.24\textwidth]{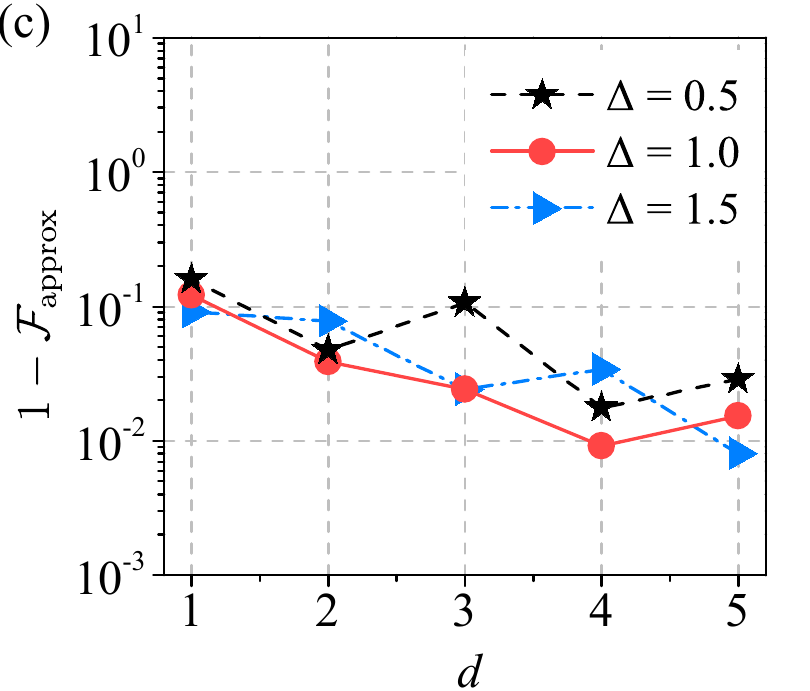}&
  \includegraphics[width=0.24\textwidth]{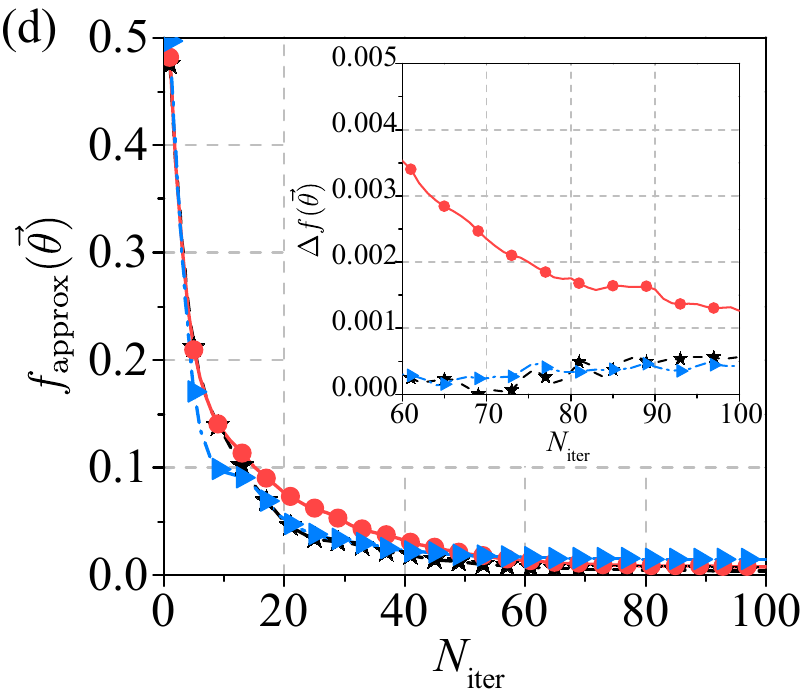}\\
  \end{tabular}
  \caption{(a) Final fidelity as a function of the circuit depth given $L=15$ for ideally computed fidelities. (b) Loss values as a function of the number of iterations given $L=15$ and $d=20$ for ideally computed fidelities. (c) Final fidelity as a function of the circuit depth given $L=6$ with fidelities computed by the SWAP test. (d) Loss values as a function of the number of iterations given $L=6$ and $d=5$, with fidelities computed by the SWAP test. The inset is the difference between $f_{\text {ideal}(\vec{\theta})}$ computed with ideal fidelities and $f_{\text {approx}(\vec{\theta})}$ evaluated via SWAP test. In all the figures, the blue dash-dotted lines with triangles, the red solid lines with circles and the black dashed lines with stars correspond to $\Delta=0.5, 1.0$ and $1.5$ respectively.}
  \label{Fig:Result}
\end{figure}

Concretely, we adapt two different configurations in our simulations. In the first configuration, we assume that the fidelity $\fidelity$ is evaluated ideally, that is we neglect the error produced in the SWAP test when estimating $\fidelity$ from a finite number of measurements.
We prepare the ground states of an XXZ chain with 15 spins as an MPS using density matrix renormalization group method~\cite{Schollwock2011} via a classical simulator, and take this state as the target state. After that, the target states are approximated with variational circuits of different depths, and the parameters are optimized through the BFGS optimizer~\cite{BFGS}. In the second configuration, we use a classical simulator, which directly stores the quantum state as a vector, and faithfully compute $\fidelity$ using SWAP test as follows. Given two $n$-qubit quantum states $\vert \psi\rangle$ and $\vert \psi^o\rangle$, we perform the following quantum circuit evolution~\cite{Buhrman2001}
\begin{align}
(H\otimes I)(\cswap)(H\otimes I) \vert 0\rangle \vert \psi\rangle \vert \psi^o\rangle,
\end{align}
where $H$ is the Hadmard gate, $I$ is the identity operator, $\cswap$ is the controlled-swap operation, and $\vert 0\rangle$ is an auxiliary qubit. Then we measure the probability that the auxiliary qubit is in state $\vert 1\rangle$, which is denoted by $p_{\vert 1\rangle}$, and is equal to $(1-|\langle \psi^o \vert \psi\rangle|^2)/2$. Thus, we have
\begin{align}
|\langle \psi^o \vert \psi\rangle| = \sqrt{1 - 2p_{\vert 1\rangle}}.
\end{align}
However, in practice, one can only perform a finite number of quantum measurements and obtain $p_{\vert 1\rangle}$ approximately. Therefore, in this configuration and also for a real quantum computer, the fidelity $\fidelity$ as well as the gradients can only be computed approximately. In our simulations, we perform 10000 measurements to evaluate each $\fidelity$, and apply the ADAM gradient-based optimizer~\cite{ADAM} to minimize the loss. The results show that the ground state can still be approximated with high precision. Due to the heavy cost to faithfully simulate the SWAP test, we only simulate up to 6 spins in the second case. The results of the simulations are summarized in Fig.~\ref{Fig:Result}.

In Fig.~\ref{Fig:Result}(a, b), we show the numerical results for $L=15$ where the fidelities are computed ideally. In Fig.~\ref{Fig:Result}(a), we plot the final fidelity between the output of the variational quantum circuits and $\vert GS(\Delta)\rangle$ as a function of the depth $d$ for $\Delta=0.5, 1$ and $1.5$ respectively. We can see that with a circuit depth of $d=15$ ($240$ parameters), the final fidelity reaches above $99.1\%$ for all the cases. For larger value of $\Delta$, it requires a fewer number of parameters to reach the same precision. This result meets our expectations since in the gapped phase, the entanglement of the ground state is bounded~\cite{Hastings2012}. In Fig.~\ref{Fig:Result}(b), we plot the loss against the number of iterations. For $\Delta=$ 0.5, 1, 1.5, reaching $f(\para) \leq 0.05$ requires $N_{\text {iter}} = $ 50, 44, 34 and reaching $f(\para) \leq 0.01$ requires $N_{\text {iter}} =$ 240, 209, 206, respectively. In Fig.~\ref{Fig:Result}(c, d), we show the numerical results for $L=6$ with $\fidelity$ faithfully computed using the SWAP test. In Fig.~\ref{Fig:Result}(c), we plot the final fidelity for $L=6$ after $100$ iterations. The state can be approximated with a fidelity beyond 95\% by using a low-depth circuit. Figure~\ref{Fig:Result}(d) shows the loss against the number of iterations. The inset shows the comparison $\Delta f(\vec{\theta})=\left|f_{\text {approx}(\vec{\theta})}-f_{\text {ideal}(\vec{\theta})}\right|$ between $f_{\text {ideal}(\vec{\theta})}$ computed with ideal fidelities and $f_{\text {approx}(\vec{\theta})}$ evaluated via SWAP test, starting from the same set of initial parameters. All the results suggest that the algorithm has a remarkable robustness even if $\fidelity$ can not be computed exactly.

\section{Conclusion}\label{sec:summary}

In this work, we proposed a method for quantum state tomography. We first utilize quantum machine learning to extract the information of the target quantum state into a VQC, which requires only a polynomial number of gate operations for a quantum computer and hopefully can be executed on near-term quantum computers. Then based on the determined variational quantum circuit, the information of the target state can be efficiently recovered using a MPS circuit simulator. Our method can be applied for the tomography of both pure states and mixed states. For pure state, the tomography can be further simplified if this constraint of target state is known in advance.
We demonstrate our method by approximating ground states of a local spin Hamiltonian with 6 or 15 qubits based on a VQC simulator. The result indicates that a high fidelity could be reached with a relatively small number of variational parameters and iterations, and this method can hopefully be applied on near-term quantum devices.

\begin{acknowledgments}

The numerical simulation is done by the open source variational quantum circuit simulator VQC~\cite{VQC}.
We gratefully acknowledge the help from China Greatwall Technology. We appreciate the helpful discussion with other members of QUANTA group. J. W. acknowledges the support from the National Natural Science Foundation of China under Grant 61632021. C. G. acknowledges the support from the National Natural Science Foundation of China under Grants No. 11805279 and No. 11504430. P. X. acknowledges support from National Natural Science Foundation of China under Grants No. 11621091 and No. 11690031. X. Q. acknowledges support from National Natural Science Foundation of China under Grants No. 11804389. A. H. acknowledges support from the National Natural Science Foundation of China under Grants No. 61901483 and the National Key Research and Development Program of China under Grants No. 2019QY0702. H.-L. H. acknowledges support from the Open Research Fund from State Key Laboratory of High Performance Computing of China (Grant No. 201901-01), National Natural Science Foundation of China under Grant No. 11905294, and China Postdoctoral Science Foundation.
\end{acknowledgments}

%

\end{document}